\documentclass[apj,iop,twocolumn,numberedappendix]{emulateapj-rtx4}
\usepackage{graphicx}
\usepackage{mathrsfs}
\usepackage[intlimits,centertags]{amsmath}
\usepackage{amssymb,amsfonts}
\usepackage{enumerate}
\usepackage[pdftex]{hyperref}
\usepackage[x11names]{xcolor}

\hypersetup{pdftitle={Maximum-Likelihood Technique for Reconstructing Cosmic-Ray Anisotropy},
pdfsubject={Maximum-Likelihood Technique for Reconstructing Cosmic-Ray Anisotropy},
pdfauthor={Markus Ahlers et al.},
pdfstartview={FitH},
colorlinks=true,
bookmarksopen=false,
bookmarksnumbered=false,
bookmarksopenlevel=0,
linkcolor=Blue1!60!black,
citecolor=Green1!50!black,
urlcolor=Blue1!70!black
}

\begin{document}

\title{A New Maximum-Likelihood Technique for\\ Reconstructing Cosmic-Ray Anisotropy at All Angular Scales}
\shorttitle{Maximum-Likelihood Technique for Reconstructing Cosmic-Ray Anisotropy}
\shortauthors{Ahlers, BenZvi, Desiati, D\'iaz--V\'elez, Fiorino \& Westerhoff}

\author{M.Ahlers\altaffilmark{1,\textdagger}}
\author{S.Y.BenZvi\altaffilmark{2}}
\author{P.Desiati\altaffilmark{1}}
\author{J.C.D\'iaz--V\'elez\altaffilmark{1,3}}
\author{D.W.Fiorino\altaffilmark{1,4}}
\author{S.Westerhoff\altaffilmark{1}}
\altaffiltext{1}{WIPAC \& Department of Physics, University of Wisconsin--Madison, Madison, WI 53706, USA}
\altaffiltext{2}{Department of Physics \& Astronomy, University of Rochester, Rochester, NY 14627, USA}
\altaffiltext{3}{Centro Universitario de los Valles, Universidad de Guadalajara, Guadalajara, Jalisco 44130, M\'exico}
\altaffiltext{4}{now at Department of Physics, University of Maryland, College Park, MD, USA}
\altaffiltext{\textdagger}{\href{mailto:mahlers@icecube.wisc.edu}{mahlers@icecube.wisc.edu}}

\submitted{}

\begin{abstract}
The arrival directions of TeV--PeV cosmic rays show weak but significant anisotropies with relative intensities at the level of one per mille. Due to the smallness of the anisotropies, quantitative studies require careful disentanglement of detector effects from the observation. We discuss an iterative maximum-likelihood reconstruction that simultaneously fits cosmic ray anisotropies and detector acceptance. The method does not rely on detector simulations and provides an optimal anisotropy reconstruction for ground-based cosmic ray observatories located in the middle latitudes. It is particularly well suited to the recovery of the dipole anisotropy, which is a crucial observable for the study of cosmic ray diffusion in our Galaxy. We also provide general analysis methods for recovering large- and small-scale anisotropies that take into account systematic effects of the observation by ground-based detectors.
\end{abstract}


\maketitle

\section{Introduction} 

During the past decade, a number of cosmic-ray, $\gamma$-ray, and neutrino observatories have found anisotropies in the arrival directions of Galactic cosmic rays at TeV and PeV energies~\citep{Guillian:2005wp,Amenomori:2006bx,Abdo:2008kr, Abdo:2008aw, Abbasi:2011ai,Abbasi:2011zka,Amenomori:2012uda,Aartsen:2013lla,ARGO-YBJ:2013gya,Abeysekara:2014sna,Bartoli:2015ysa}. For a recent summary of experimental results, we refer to~\cite{DiSciascio:2014jwa}. The statistics and resolution of these experiments allow for a two-dimensional reconstruction in the form of anisotropy sky maps. The studies have revealed significant anisotropies at both large and small angular scales.  At the largest scales, the anisotropy is approximately dipolar and has a relative intensity on the order of $10^{-3}$. An explanation of the amplitude and phase of the dipole anisotropy is challenging, but the observations are qualitatively consistent with diffusive propagation of cosmic rays from Galactic sources~\citep{Ptuskin2006,Erlykin:2006ri,Blasi:2011fm,Mertsch:2014cua}. 

Medium- and small-scale anisotropies have been observed at the $10^{-4}$ level. These features are less understood but could be a combined effect of nearby cosmic-ray sources~\citep{Salvati:2008dx, Biermann:2012tc} and the local interstellar magnetic field structure, which can introduce an energy-dependent magnetic mirror leakage~\citep{Drury:2008ns}, preferred cosmic-ray transport directions~\citep{Malkov:2010yq}, or magnetic lenses~\citep{Battaner2011, Battaner:2014cza}. The observed power spectrum of cosmic-ray anisotropies agrees well with the expected effect of cosmic-ray scattering in the local magnetic field~\citep{Giacinti:2011mz, Ahlers:2013ima, Ahlers:2015dwa, Lopez-Barquero:2015qpa}.  The possible influence of the heliosphere via magnetic reconnections in the heliotail~\citep{Lazarian:2010sq}, non-isotropic particle transport in the heliosheath~\citep{Desiati:2011xg} or the heliospheric electric field structure~\citep{Drury:2013uka} has also been considered. More exotic origin models invoke strangelet production in molecular clouds~\citep{Kotera:2013mpa} or in neutron stars~\citep{Perez-Garcia:2013lza}.

\begin{figure}[t]
\centering
\includegraphics[width=\linewidth]{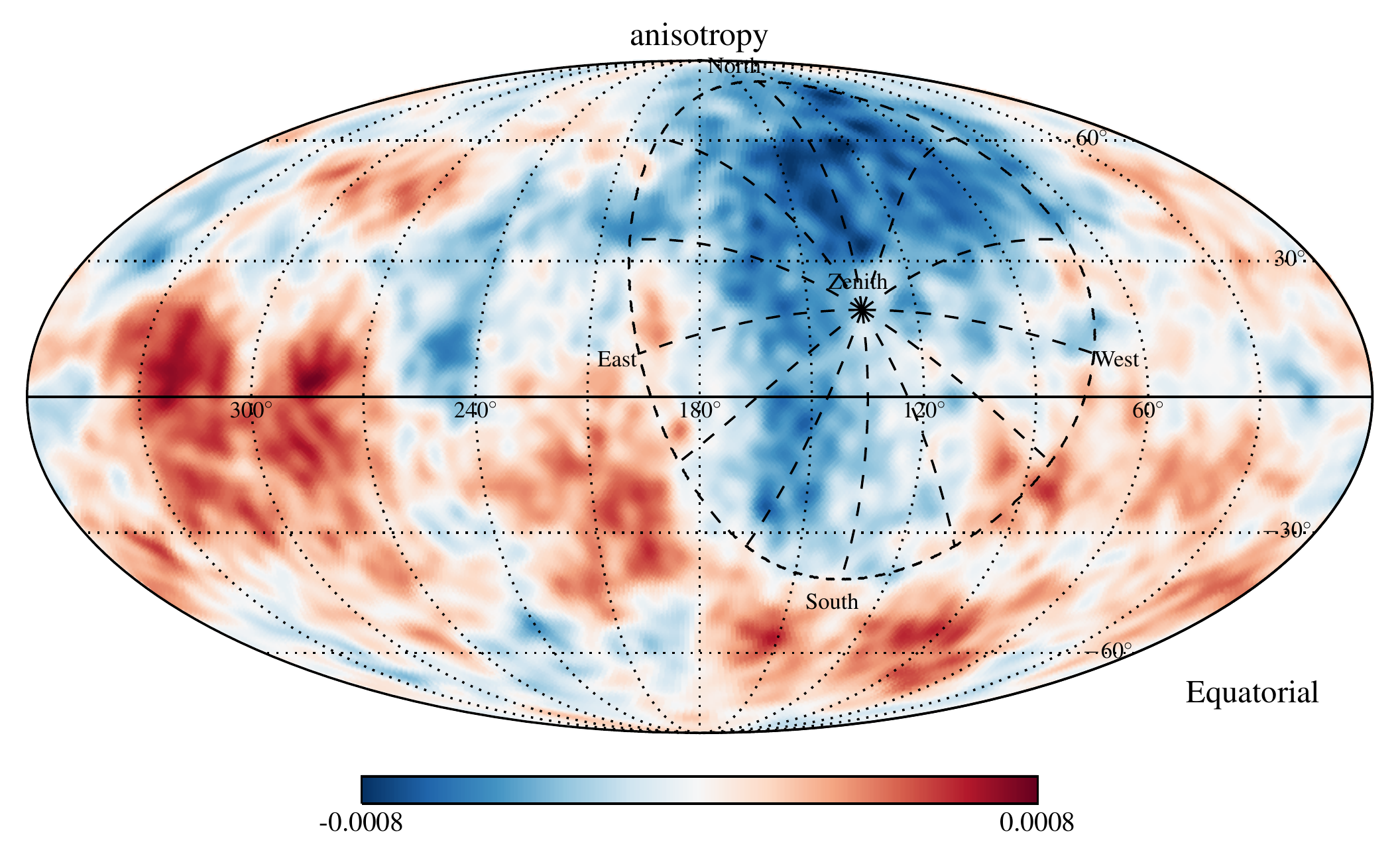}
\caption[]{Simulated cosmic ray anisotropy in equatorial coordinates using the model of \cite{Ahlers:2013ima}. For illustration, we indicate the instantaneous field of view of the HAWC observatory (at latitude $19^\circ$ north) at a local sidereal time of 9h and a zenith angle cut at $60^\circ$. The time-integrated field of view corresponds to the declination range $-41^\circ<\delta<79^\circ$.}\label{fig1}
\end{figure}

The measurement of cosmic-ray anisotropies with relative intensity below $10^{-3}$ is an observational challenge, since it is necessary to account for minuscule variations in the acceptance and uptime of the detector carrying out the measurement. For illustration, Fig.~\ref{fig1} shows a simulated realization of the cosmic-ray anisotropy following the model of~\cite{Ahlers:2013ima}. This model was chosen due to its qualitative agreement with the observation of anisotropy in TeV cosmic rays by IceCube~\citep{Aartsen:2013lla} and HAWC~\citep{Abeysekara:2014sna}. The cosmic-ray anisotropy is expected to remain constant in the celestial coordinate system over the period of the observation.  Ground-based cosmic observatories with a limited field of view, however, are exposed to different parts of the celestial sphere as the Earth rotates during one sidereal day. As an example, the sky map of Fig.~\ref{fig1} indicates the instantaneous field of view of the HAWC detector~\citep{Abeysekara:2014sna} (at latitude $19^\circ$~N) at a local sidereal time of 9h. Hence, the observed event distribution accumulated over many sidereal days depends not only on the cosmic-ray anisotropy but also on the nonuniform and time-dependent detector exposure. It is therefore necessary to reconstruct a reference map that represents the response of the detector to an isotropic cosmic-ray flux. For ground-based detectors, in which the atmosphere acts as part of the observatory, sufficiently accurate simulations of the detector exposure are usually not achievable. Therefore, most analyses use the data themselves to estimate the relative intensity and detector exposure simultaneously. 

Examples of this technique are the {\it time-scrambling}~\citep{Alexandreas1993} and {\it direct-integration}~\citep{Atkins:2003ep} methods, in which the rate of events observed in a detector as a function of local sidereal time is integrated against the relative acceptance of the detector during an integration period (or scrambling interval) $\Delta t$. The idea behind this method is that variations in the event rates introduced by the cosmic-ray anisotropies will average out to some extent as the detector observes different parts of the celestial sphere over the course of each sidereal day. The result is an estimate of the number of events expected in the detector between time $t$ and $t+\Delta t$. Subtracting the expected number of events from the actual observations yields a residual counts map which can be explored for anisotropy.  

The integration time interval $\Delta t$ acts as an effective smoothing parameter for the counts map, since the method averages cosmic-ray arrival directions over angular scales of $15^\circ(\Delta t/1{\rm hr})$. In principle, choosing $\Delta t=24$h would produce a residual counts map with features covering the full sky ($360^\circ$). However, for detectors located in the middle latitudes, the instantaneous exposure of the detector does not match the full daily exposure, since the 24h integrated field of view is much larger than the instantaneous field of view, {cf.}~Fig~\ref{fig1}. As a result, large-scale structures in the residual counts map, in particular the dipole, are strongly attenuated when using these methods. 

To improve estimates of large-scale anisotropy using detectors in the middle latitudes, we describe a maximum likelihood construction that can be used to disentangle the anisotropy from detector effects. The technique is based on the same ansatz used by the time-scrambling or direct-integration method, that the total accumulated exposure of the detector can be factorized into a time-dependent event rate and a time-independent relative acceptance map. We begin by describing the technique in Section~\ref{sec:lratio}. In Section~\ref{sec:simulation}, we apply the maximum-likelihood method to simulated data and show that large- and small-scale anisotropies can be reconstructed with relatively little of the distortion observed in the direct-integration or time-scrambling techniques. We compare our method to alternative techniques in Section~\ref{sec:comparison}. We then discuss analysis methods of large- and small-scale anisotropies of the reconstructed anisotropy maps in Section~\ref{sec:harmonic} before concluding in Section~\ref{sec:summary}.

\section{Maximum Likelihood Method}\label{sec:lratio}
 
In the following, we will assume that the total accumulated detector exposure $\mathcal{E}$ can be expressed as a product of its angular-integrated exposure $E$ and relative acceptance $\mathcal{A}$ in terms of azimuth angle $\varphi$ (from north increasing to the east) and zenith angle $\theta$ as
\begin{equation}\label{eq:E}
  \mathcal{E}(t,\varphi,\theta) \simeq E(t)\mathcal{A}(\varphi,\theta)\,.
\end{equation}
Without loss of generality, we require that the relative acceptance is normalized to $\int{\rm d}\Omega \mathcal{A}(\varphi,\theta)=1$. This approximation assumes that the relative acceptance of the detector remains approximately constant over time. The ansatz is identical to the approach used in direct integration or time scrambling.

Let us also assume that the flux of cosmic rays at the energies of interest remains constant as a function of time, varying only as a function of celestial longitude $\alpha$ (right ascension) and latitude $\delta$ (declination).  Due to the strong diffusion of cosmic rays in the Galactic environment, the flux is dominated by an isotropic term $\phi^{\rm iso}$. Hence, the total flux can be expressed as 
\begin{equation}\label{eq:phi}
  \phi(\alpha,\delta) = \phi^{\rm iso}I(\alpha,\delta)\,,
\end{equation} 
where $I(\alpha,\delta)$ is the {\it relative intensity} of the flux as a function of position in the sky. The {\it anisotropy} is defined as the deviation $\delta I = I-1\ll 1$. Note that this ansatz ignores anisotropies associated with the relative motion of the Earth with respect to the Sun. We will come back to this subtlety in the discussion section. 

The local horizontal coordinate system and the celestial (or equatorial) coordinate system are related via a time-dependent transformation. We define ${\bf n}=(\cos\alpha\cos\delta,\sin\alpha\cos\delta,\sin{\delta})$ as the unit vector corresponding to the coordinates $(\alpha,\delta)$ in the right-handed equatorial system.  Similarly, the unit vector corresponding to the coordinates $(\theta,\varphi)$ in the right-handed local system is ${\bf n}'=(\cos\varphi\sin\theta,-\sin\varphi\sin\theta,\cos{\theta})$. The two unit vectors are related via a time-dependent coordinate transformation ${\bf n}'={\bf R}(t){\bf n}$. For an experiment located at geographic latitude $\Phi$ and longitude $\Lambda$ (measured east from Greenwich), the transformation is
\begin{equation}
{\bf R}(t) =
\begin{pmatrix}
  -\cos \omega t\sin \Phi&-\sin \omega t\sin\Phi&\cos\Phi \\
  \sin \omega t&-\cos \omega t&0\\
  \cos \omega t\cos\Phi&\sin \omega t\cos\Phi&\sin\Phi
\end{pmatrix}\,,
\end{equation}
where $\omega = 2\pi/24$h and the local sidereal time $t$ is related to the sidereal time at Greenwich $t'$ by $t=t'+\Lambda/\omega$. 

To simplify calculations on the local and celestial spheres, the sky is binned  into pixels of equal area $\Delta\Omega$ using the {\tt HEALPix} parametrization of the unit sphere~\citep{Gorski:2004by}. To make the equations more transparent, we use {\it roman} indices for pixels in the local sky map and {\it fraktur} indices for pixels in the celestial sky map. Time bins are indicated by {\it greek} indices. For instance, the data observed at a fixed sidereal time bin $\tau$ can be described in terms of the observation in local horizontal sky with bin $i$ as $n_{\tau i}$ or transformed into the celestial sky map with bin $\mathfrak{a}$ as $n_{\tau \mathfrak{a}}$. 

Consider an angular element of the local coordinate sphere $\Delta\Omega_i$ corresponding to coordinates $(\theta_i,\varphi_i)$. The number of cosmic rays expected from this location in a sidereal time interval $\Delta t_\tau$ with central value $t_\tau$ is
\begin{equation}\label{eq:mu}
  \mu_{\tau i} \simeq I_{\tau i}\mathcal{N}_\tau\mathcal{A}_{i}\,,
\end{equation}
where $\mathcal{N}_\tau\equiv\Delta t_\tau\phi^{\rm iso}{E}(t_\tau)$ gives the expected number of isotropic background events in sidereal time bin $\tau$. The quantity $\mathcal{A}_i \equiv \Delta\Omega_i\mathcal{A}(\theta_i,\varphi_i)$ is the binned relative acceptance of the detector for angular element $i$, and $I_{\tau i}\equiv {I}({\bf R}(t_\tau){\bf n}'(\Omega_\mathfrak{i}))$ is the relative intensity observed in the local horizontal system during time bin $\tau$.  For simplicity, all expressions that follow assume equal bin sizes $\Delta\Omega=4\pi/N_{\rm pix}$ on the local and celestial spheres and $N_{\rm time}$ equal sidereal bins with $\Delta t = 24{\rm h}/N_{\rm time}$. However, a more general binning is also possible.

Given $\mu_{\tau i}$, the likelihood of observing $n$ cosmic rays is given by the product of Poisson probabilities
\begin{equation}\label{eq:LH}
  \mathcal{L}(n|I,\mathcal{N},\mathcal{A}) =
  \prod_{\tau i}\frac{(\mu_{\tau i})^{n_{\tau i}}e^{-\mu_{\tau i}}}{n_{\tau i}!}\,,
\end{equation}
where $n_{\tau i}$ is the number of events observed in the local pixel $i$ during time bin $\tau$. This likelihood can be maximized to provide estimators of the relative acceptance function $\mathcal{A}_i$ and the expected isotropic background count $\mathcal{N}_\tau$.

Consider the case of the null hypothesis of no anisotropy, which we denote $I_\mathfrak{a}^{(0)}=1$. Given the boundary condition $\sum_i\mathcal{A}_i=1$, the maximum likelihood estimators of $\mathcal{A}_i$ and $\mathcal{N}_\tau$ are
\begin{align}\label{eq:Nnull}
  \mathcal{N}_\tau^{(0)} &=  {\sum_i n_{\tau i}}\,,\\\label{eq:Anull}
  {\mathcal{A}}_i^{(0)} &= \sum_\tau n_{\tau i}\Big/\sum_{\kappa j}n_{\kappa j}\,.
\end{align}
To allow for the possibility of anisotropy, we maximize the likelihood ratio
\begin{equation}\label{eq:LHratio}
  \lambda = \frac{\mathcal{L}(n|I,\mathcal{N},\mathcal{A})}
                 {\mathcal{L}(n|I^{(0)},\mathcal{N}^{(0)},\mathcal{A}^{(0)})}
\end{equation}
of signal over null hypothesis in $\mathcal{N}$, $\mathcal{A}$, and $I$.

\subsection{Invariance under Declination-Scaling}

We will demonstrate in the following that the maximum-likelihood method can be used to search for anisotropy on all angular scales. However, before we continue, note that events recorded in a fixed position in the local coordinate system can only probe the cosmic-ray flux in a fixed declination band $\delta$. Hence, the expectation values (\ref{eq:mu}) are invariant under the rescaling
\begin{eqnarray}
\label{eq:scaleI}
  I \to & I' &\equiv I/a(\delta)/b\,,\\\label{eq:scaleN}
  \mathcal{N}\to &{\mathcal{N}}'&\equiv \mathcal{N}bc
\,,\\\label{eq:scaleA}
  \mathcal{A} \to &{\mathcal{A}}'&\equiv{\mathcal{A}a(\delta)}/{c}\,,\end{eqnarray}
where $a(\delta)$ is an arbitrary function of declination and the normalization factors $b$ and $c$ are defined such that $\sum_\tau{\mathcal{A}_\tau}'=1$ and $\sum_\mathfrak{a} \delta I'_\mathfrak{a}=0$ for the new values. In other words, the maximum-likelihood method is sensitive to anisotropy in right ascension but is insensitive to variations in intensity across declination bands. This is also known to be a limitation of other reconstruction methods, like direct integration or time scrambling~\citep{Amenomori:2005pn,Iuppa:2013pg}. Because of this degeneracy, we must choose whether to account for the local excess of cosmic rays as originating in an anisotropy in their relative intensity or originating in a variation in the local acceptance.

A natural choice is that the anisotropy is normalized to $\int{\rm d}\alpha\delta I(\alpha,\delta)=0$ for all declinations $\delta$, consistent with the definition $\int{\rm d}\Omega \delta I(\alpha,\delta)=0$. This condition can also be formulated in terms of a spherical harmonics expansion of the relative intensity in the equatorial coordinate system $(\alpha,\delta)$, as pointed out by~\citet{Iuppa:2013pg}. In general, the relative intensity can be decomposed as a sum over spherical harmonics $Y^{\ell m}$,
\begin{equation}\label{eq:Ylm}
{\delta I}_{\mathfrak{a}} = \sum_{\ell\geq1}\sum_{m=-\ell}^\ell \widehat{a}_{\ell m} Y^{\ell m}_{\mathfrak{a}}\,.
\end{equation}
Our normalization condition can then be expressed as the condition $\widehat{a}_{\ell 0}=0$ for all $\ell$. This projection significantly reduces the reconstruction of the low-$\ell$ multipole components of the anisotropy, as we will discuss in the following sections.

Note that the true multipole moments $\widehat{a}_{\ell m}$ are an (infinite) superposition of the {\it pseudo} multipole moments ${a}_{\ell m}$, which are defined as in Eq.~(\ref{eq:Ylm}), but for the product of the relative intensity with the weight function $w$ of the field of view. Provided that the weight function is azimuthally symmetric, $w(\alpha,\delta) = w(\delta)$, the true multipole moments $\widehat{a}_{\ell 0}$ are a linear superposition of pseudo multipole moments ${a}_{\ell' 0}$. In practice, we can hence use the normalization condition ${a}_{\ell 0} = 0$ for all $\ell$ to ensure $\widehat{a}_{\ell 0} = 0$ for all $\ell$. In terms of the binned relative intensity and weight function this is equivalent to the condition $\sum_\mathfrak{a}w_\mathfrak{a}Y^{\ell 0}_\mathfrak{a}\delta I_\mathfrak{a} =0$ for all $\ell$.
 
\subsection{Maximum Likelihood Algorithm} 

The maximum
$(I^\star,\mathcal{N}^\star,\mathcal{A}^\star)$ of the likelihood ratio (\ref{eq:LHratio}) must obey the implicit relations
\begin{align}\label{eq:Istar}
  {I}^\star_{\mathfrak{a}} &=
  \sum_{\tau} n_{\tau\mathfrak{a}}\Big/ \sum_{\kappa}\mathcal{A}^{\star}_{\kappa \mathfrak{a}}\mathcal{N}^{\star}_\kappa\,, \\\label{eq:Nstar}
  \mathcal{N}^{\star}_\tau &=
  \sum_{i}  n_{\tau i}\Big/\sum_{j}\mathcal{A}^{\star}_jI^\star_{\tau j}\,,\\\label{eq:Astar}
\mathcal{A}^{\star}_i&= \sum_\tau n_{\tau i}\Big/\sum_{\kappa}\mathcal{N}^{\star}_\kappa I^\star_{\kappa i}\,,
\end{align}
together with $\sum_\mathfrak{a}w_\mathfrak{a}Y^{\ell 0}_\mathfrak{a}\delta I^\star_\mathfrak{a} =0$ and $\sum_i \mathcal{A}^{\star}_i =1$.  In Eq.~(\ref{eq:Istar}), we introduced the
binned quantity $\mathcal{A}_{\tau\mathfrak{a}}\equiv \Delta\Omega_\mathfrak{a}\mathcal{A}({\bf R}^T(t_\tau){\bf n}(\Omega_\mathfrak{a}))$ corresponding to the relative acceptance seen in the equatorial coordinate system in pixel $\mathfrak{a}$ during time bin $\tau$.

Equations~\eqref{eq:Istar} to \eqref{eq:Astar} correspond to a nonlinear set of equations that cannot be solved in an explicit form. However, one can approach the best-fit solution via the following iterative method:
\begin{enumerate}[\rm (i)]
\item Initialize at the maximum of the null hypothesis, $({I}^{(0)},\mathcal{N}^{(0)},\mathcal{A}^{(0)})$. 
\item Evaluate ${I}^{(n+1)}$ by inserting $({I}^{(n)},\mathcal{N}^{(n)},\mathcal{A}^{(n)})$ into the right-hand side of Eq.~(\ref{eq:Istar}).
\item Remove the $m=0$ (pseudo) multipole moments of $\delta{I}^{(n+1)}$, {i.e.}, in the equatorial coordinate system $\sum_\mathfrak{a}w_\mathfrak{a}Y^{\ell 0}_\mathfrak{a}\delta {I}_\mathfrak{a}^{(n+1)} \to 0$.
\item Evaluate $\mathcal{N}^{(n+1)}$ by inserting $({I}^{(n+1)},\mathcal{N}^{(n)},\mathcal{A}^{(n)})$ into the right-hand side of Eq.~(\ref{eq:Nstar}).
\item Evaluate $\mathcal{A}^{(n+1)}$ by inserting $({I}^{(n+1)},\mathcal{N}^{(n+1)},\mathcal{A}^{(n)})$ into the right-hand side of Eq.~(\ref{eq:Astar}).
\item Renormalize $\mathcal{N}^{(n+1)}$ and $\mathcal{A}^{(n+1)}$ as
$\mathcal{N}^{(n+1)} \to \mathcal{N}^{(n+1)}c$ and 
$\mathcal{A}^{(n+1)} \to \mathcal{A}^{(n+1)}/c$
with normalization factor $c = \sum_i\mathcal{A}_i^{(n+1)}$.
\item Repeat from step (ii) until the solution has sufficient
convergence, {i.e.}, the likelihood ratio in Eq.~\eqref{eq:LHratio} has $\Delta \chi^2 \simeq 2\ln(\lambda^{(n+1)}/\lambda^{(n)})\ll 1$.
\end{enumerate}

Note that the cosmic-ray anisotropy obtained in the first iteration step, $\delta I^{(1)}$, corresponds to the result that would be obtained by the method of direct integration~\citep{Atkins:2003ep}. This method was also used in the recent HAWC analysis~\citep{Abeysekara:2014sna}. The successive iteration steps of the maximum-likelihood method reoptimize the relative acceptance $\mathcal{A}$ and the isotropic background rate $\mathcal{N}$ for a given anisotropy. We will study this optimization process more quantitatively in the following section.

\section{Simulation and Performance}\label{sec:simulation}

To demonstrate the reconstruction of a cosmic-ray anisotropy with the maximum likelihood technique, we simulated a set of arrival directions based on the anisotropy shown in Fig.~\ref{fig1}. The simulation is based on a random realization of a relative intensity $\delta I$, which follows a power-law spectrum with $\ell\geq1$ of the form $C_\ell = 10^{-7}(18/(2\ell+1)/(\ell+1)/(\ell+2))$, after the model of \citet{Ahlers:2013ima}. This model was chosen due to its qualitative agreement with the observation of anisotropy in TeV cosmic rays by IceCube~\citep{Aartsen:2013lla} and HAWC~\citep{Abeysekara:2014sna}. The celestial sphere is binned following the {\tt HEALPix} parametrization~\citep{Gorski:2004by} with parameter $n_{\rm side}=64$ corresponding to $N_{\rm pix}=49,152$ pixels with a binsize of about $1^\circ$ diameter.

The simulated detector is located at geographic coordinates ($\Phi=19^\circ\text{N}$ and $\Lambda=97^\circ\text{W}$), the location of the HAWC observatory \citep{Abeysekara:2014sna}. The instantaneous field of view is restricted to zenith angles below $60^\circ$. A projection of the instantaneous field of view at 9h local sidereal time onto the equatorial coordinate system was already shown in Fig.~\ref{fig1}. We assume that the relative detector acceptance at any time follows $\mathcal{A}(\theta,\varphi) \propto \cos\theta[1+A\sin\theta\sin^2(\varphi - \varphi_0)]$ with $A=0.2$ and $\varphi_0=10^\circ$. The local acceptance maps are reconstructed with the same resolution as the anisotropy map.   

\begin{figure}[t]
\centering
\includegraphics[width=\linewidth]{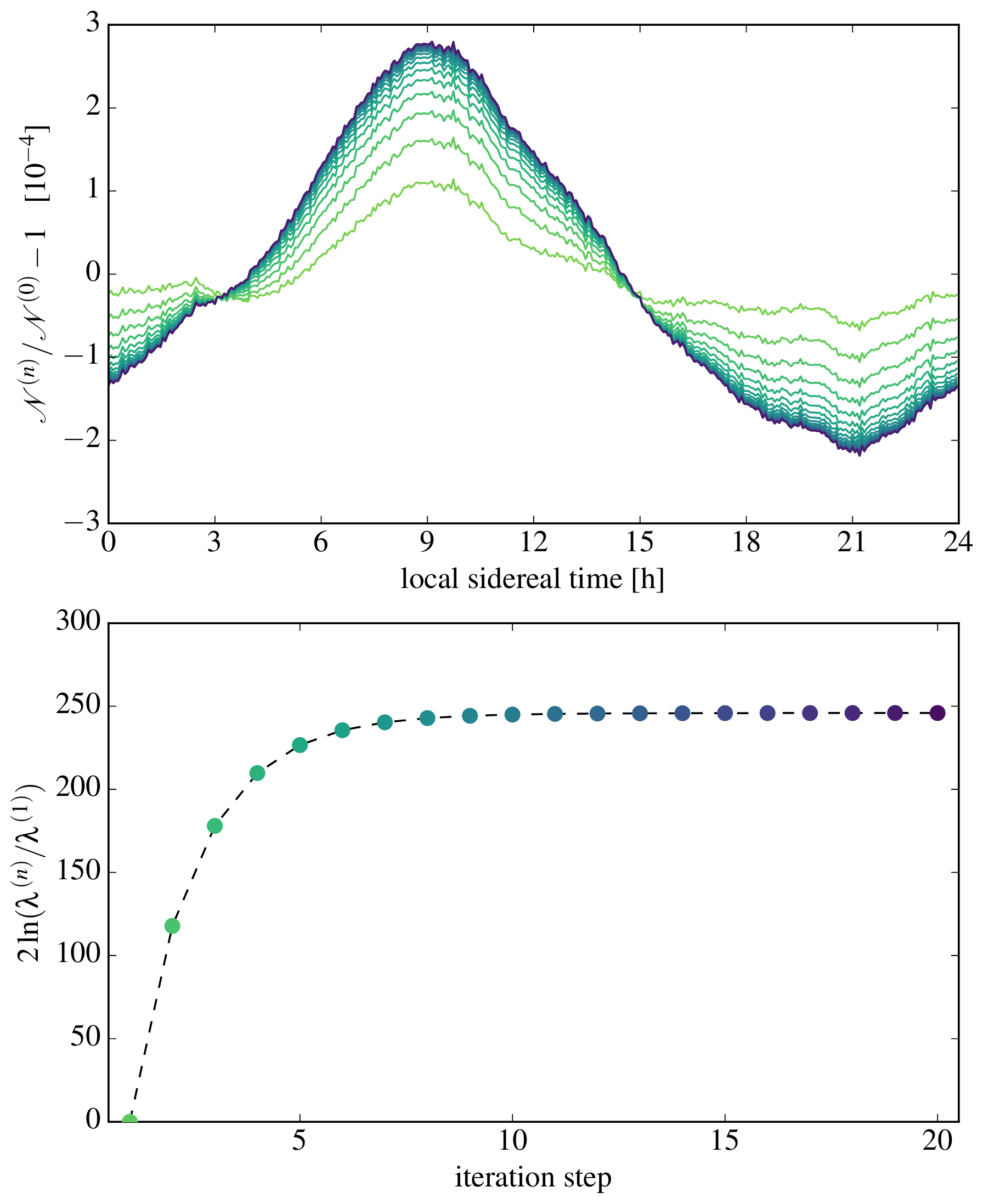}
\caption[]{
{\bf Top panel:} The relative optimization of the isotropic expectation value in terms of $\mathcal{N}^{(n)}/\mathcal{N}^{(0)}-1$ for 20 iteration steps (light to dark colors). {\bf Bottom panel:} The progressive log-likelihood values of the iteration. Note that the method already converges after about 10 iteration steps.}\label{fig2}
\end{figure}

\begin{figure*}[t]
\begin{center}
\begin{minipage}[b]{0.5\linewidth}\centering
{\scriptsize(a)}\\
\includegraphics[width=\linewidth]{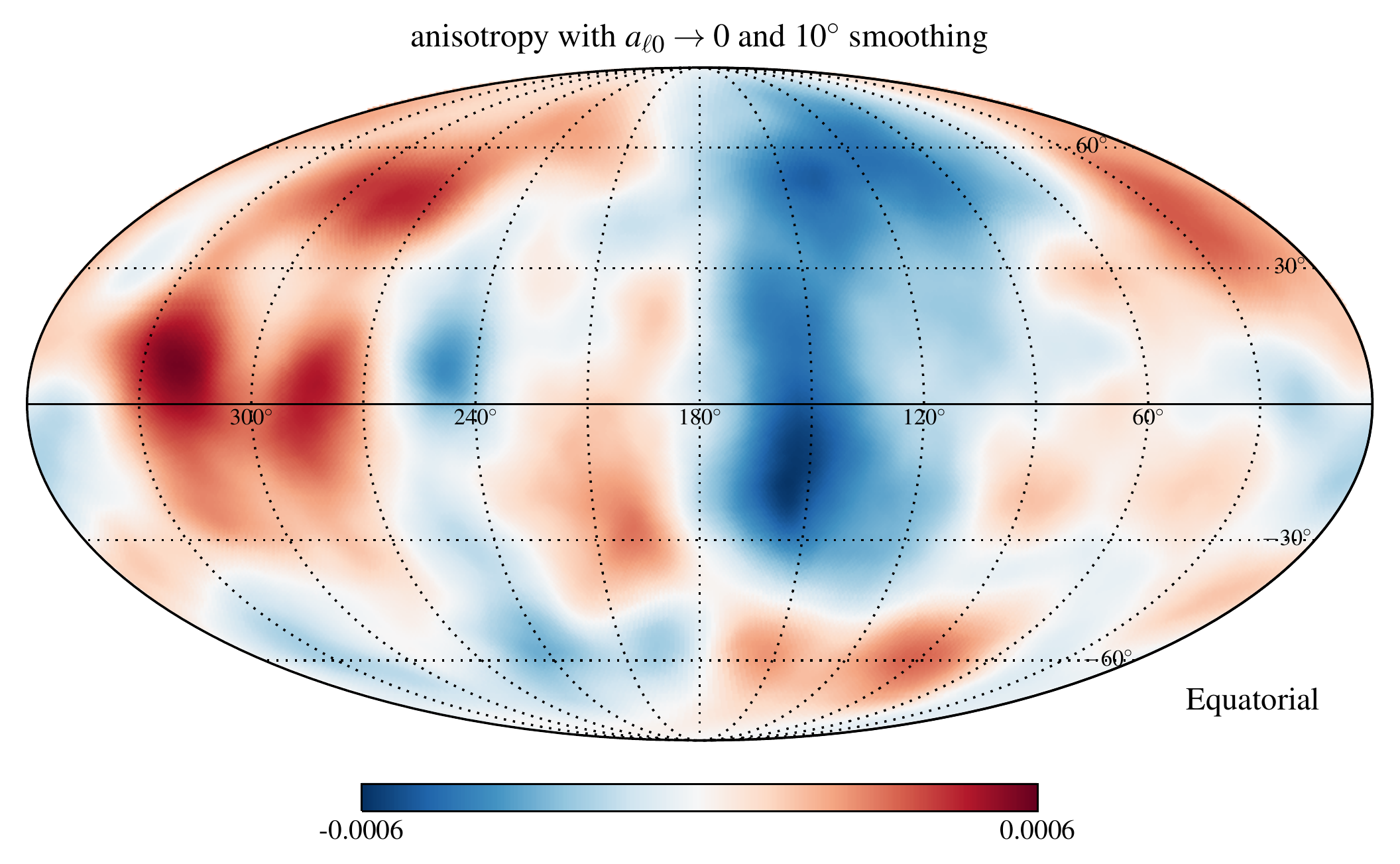}
\end{minipage}
\vspace{-0.2cm}
\end{center}
\begin{minipage}[b]{0.5\linewidth}\centering
{\scriptsize(b)}\\
\includegraphics[width=\linewidth]{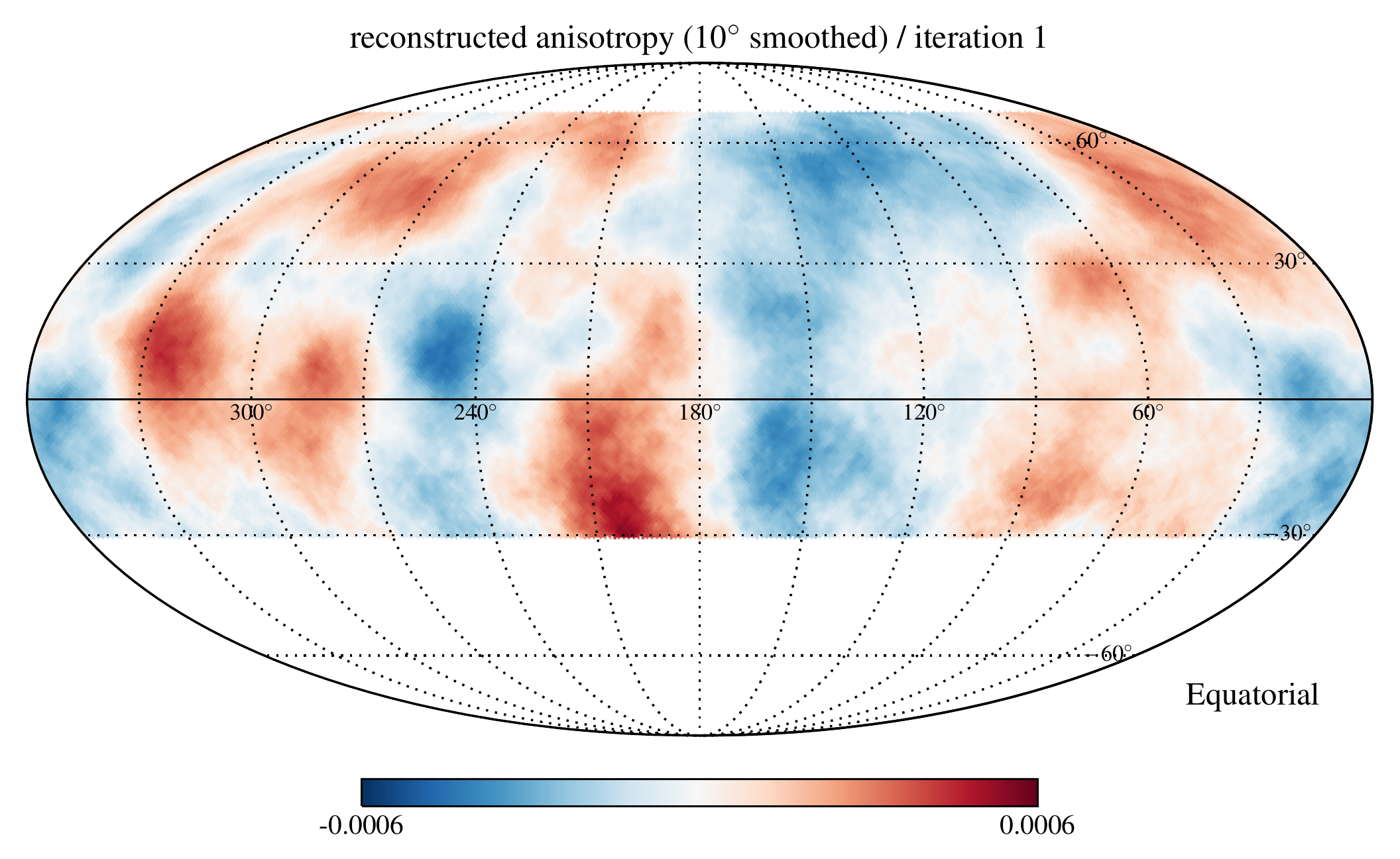}\\
{\scriptsize(d)}\\
\includegraphics[width=\linewidth]{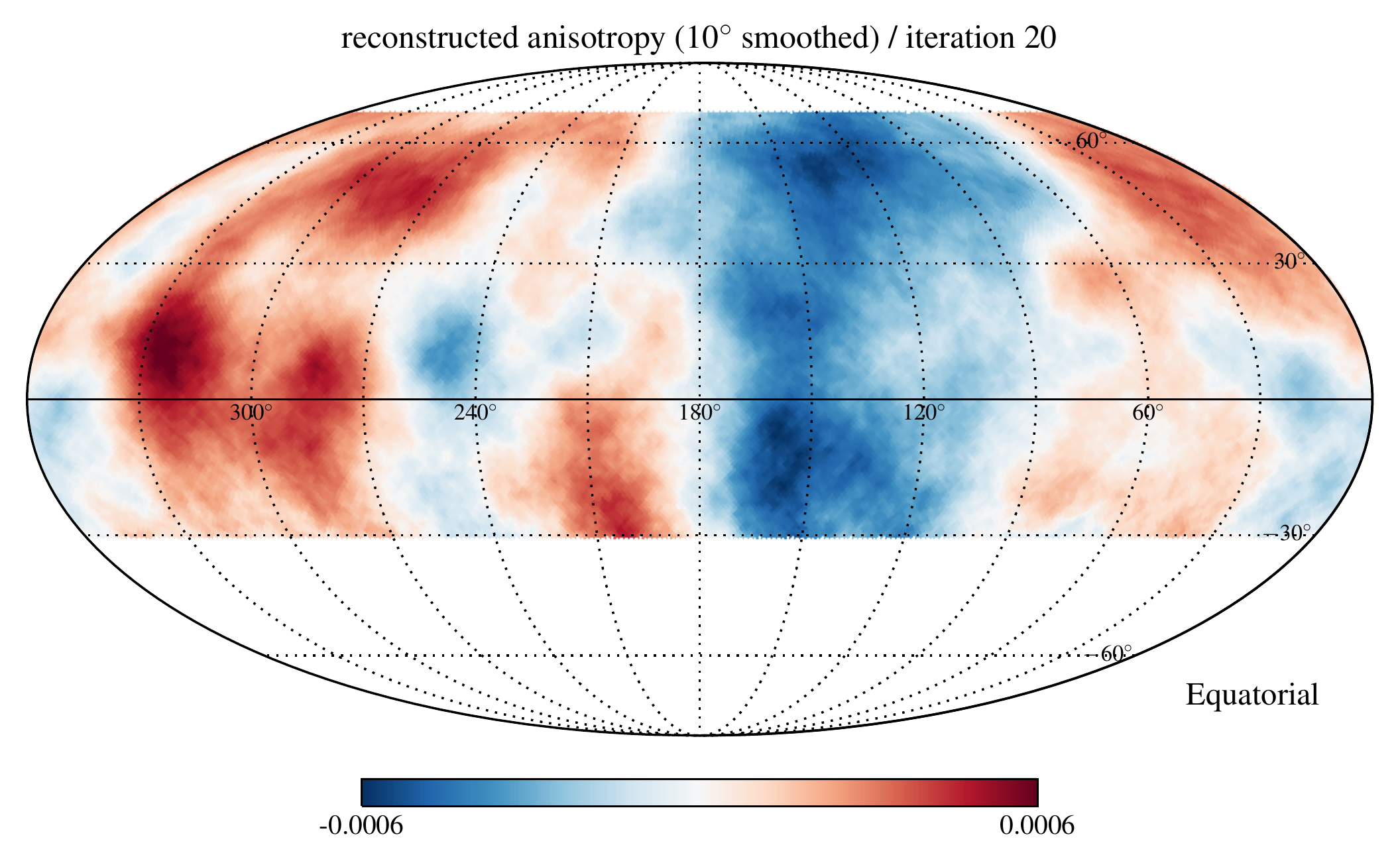}
\end{minipage}
\begin{minipage}[b]{0.5\linewidth}\centering
{\scriptsize(c) $=$ (b)$-$(a)}\\
\includegraphics[width=\linewidth]{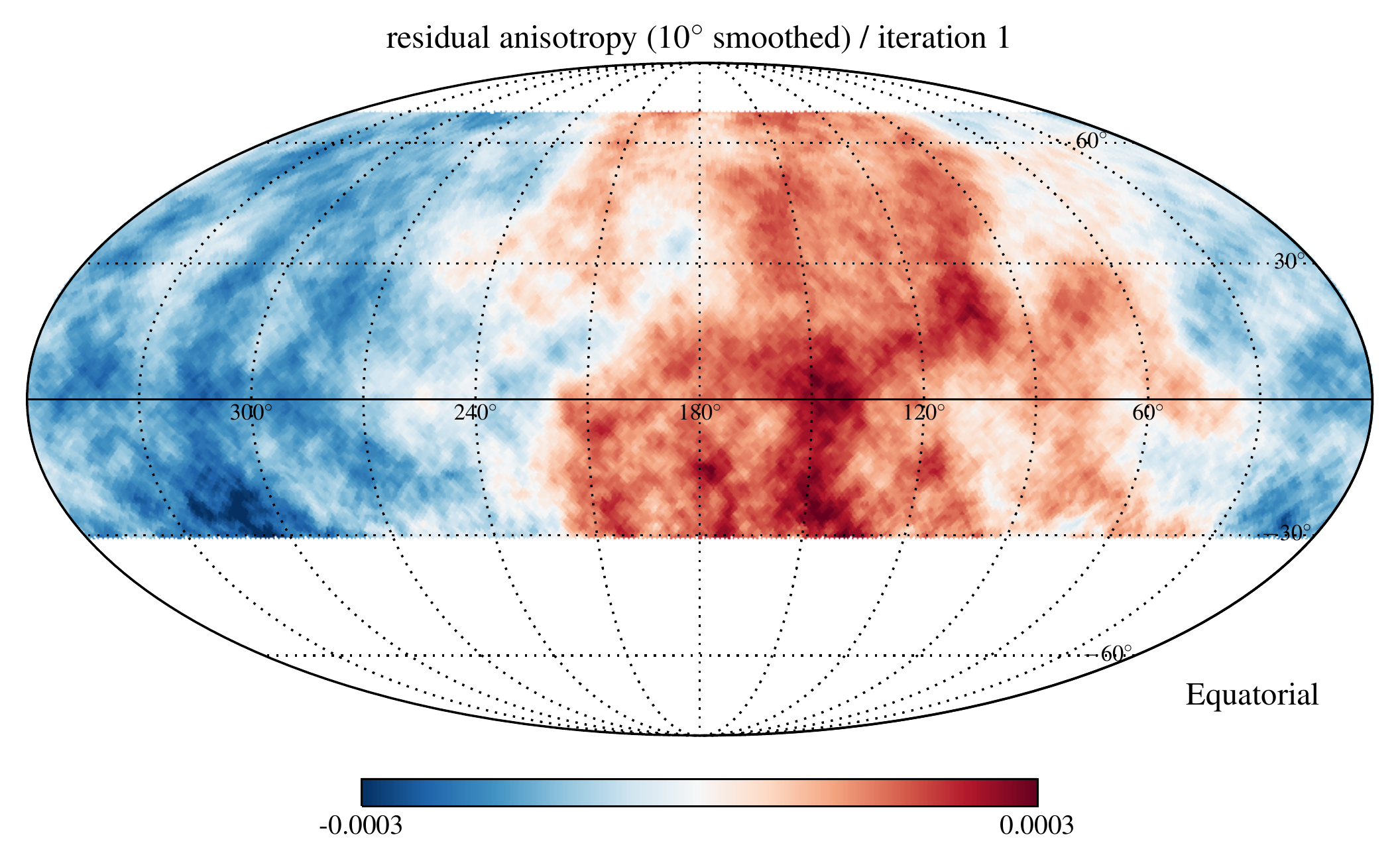}\\
{\scriptsize(e) $=$ (d)$-$(a)}\\
\includegraphics[width=\linewidth]{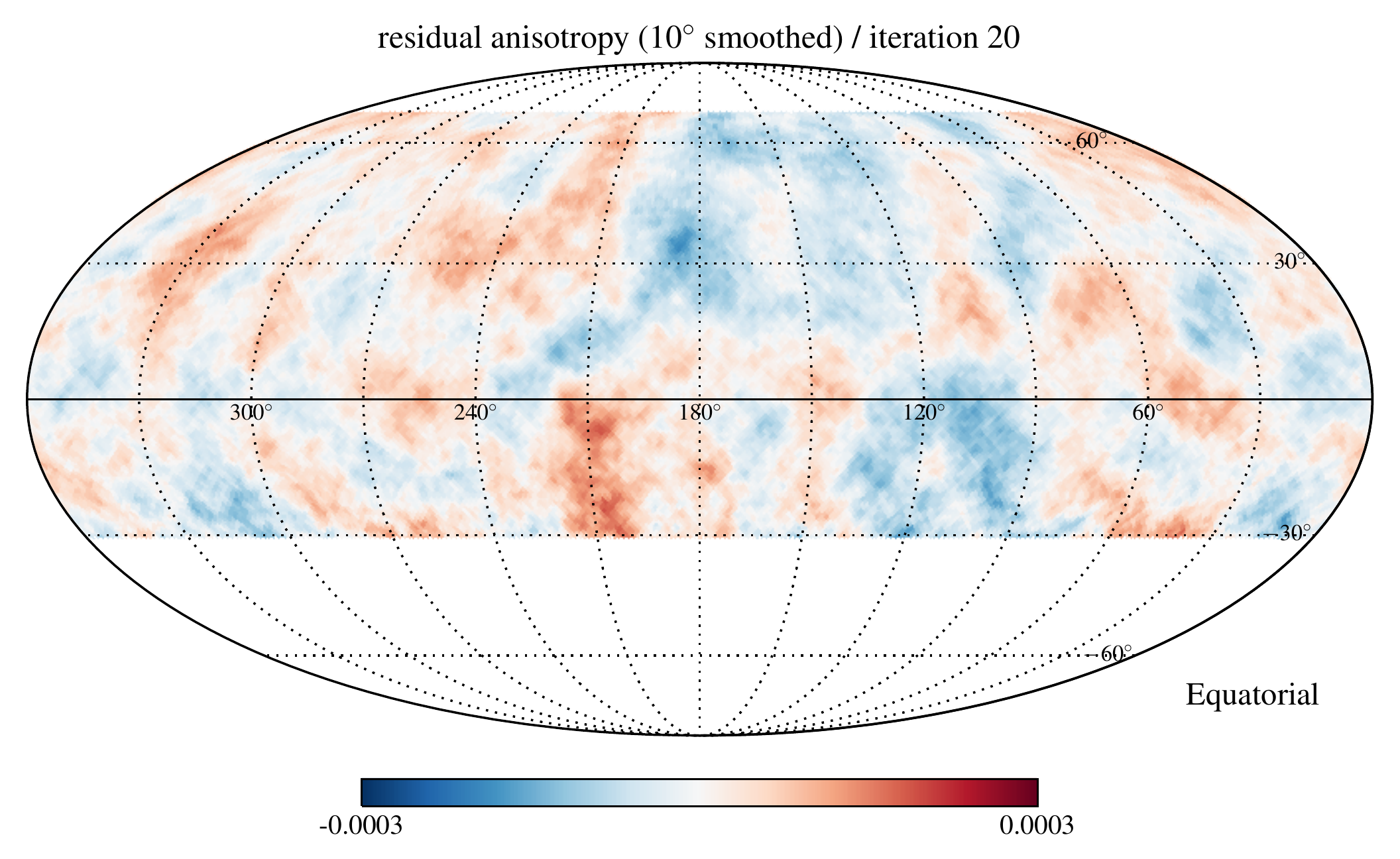}
\end{minipage}\caption[]{
{\bf Panel (a):} Expected relative intensity with $m=0$ multipole filter (see text) and $10^\circ$ smoothing.
{\bf Panel (b/d):} Reconstructed relative intensity for the first and last iteration step, respectively.
{\bf Panel (c/e):} Residual anisotropy after subtracting the expected anisotropy in panel (a) from the reconstructed anisotropy in panels (b) and (d), respectively.}\label{fig3}
\end{figure*}

The expected isotropic event number is binned in $N_{\rm time}=360$ sidereal time bins with $4$~min bin size and normalized such that $\sum_\tau\mathcal{N}_\tau\simeq5\times10^{10}$, corresponding to the integrated event number of the recent HAWC analysis~\citep{Abeysekara:2014sna}. We introduce a simple statistical toy model for the variation of the expected background rate $\mathcal{N}_\tau$; the accumulated data are expected to be stable over periods sampled from an exponential distribution with expectation value of 20~min (5 sidereal time bins). In each stable period, we assume that $\mathcal{N}_\tau$ has a normal fluctuation of 1\% around the background expectation. This model mimics the data variation of the analysis by~\cite{Abeysekara:2014sna}.

The results of the reconstruction\footnote{We implemented the iterative algorithm as a {\tt Python} program, which we can provide to the interested reader upon request.} are shown in Figs.~\ref{fig2} and \ref{fig3}. The top panel of Fig.~\ref{fig2} shows the optimization of the reconstructed background expectation $\mathcal{N}^{(n)}$ as the relative quantity $\mathcal{N}^{(n)}/\mathcal{N}^{(0)}-1$. The iteration renormalizes $\mathcal{N}^{(n)}$ at the level of $10^{-4}$ depending on the local sidereal time. The relative features introduced by the optimization process are easy to understand. For instance, the bump at 9h local sidereal time corresponds to the instantaneous field of view indicated in Fig.~\ref{fig1}. In comparison with the simulated anisotropy shown in Fig.~\ref{fig1}, one can notice that at this time the detector observes a part of the sky with a strong underfluctuation. The initial estimate $\mathcal{N}^{(0)}$ is therefore too low and the iteration compensates for this effect. The bottom panel of Fig.~\ref{fig2} shows the consecutive log-likelihood values of the iteration. Note that for this simulated data set the iteration only requires about 10 steps to converge. We run the reconstruction for 10 more iteration steps to verify the convergence and stability of the method.

The map in Fig.~\ref{fig3}a shows the expected relative intensity smoothed with a top-hat function with $10^\circ$ radius after application of the $m=0$ filter. The maps in Figs.~\ref{fig3}b and \ref{fig3}d show the reconstructed anisotropy in the first iteration step (corresponding to the result from direct integration) and after 20 iterations, respectively. Comparing the maps in Figs.~\ref{fig3}b and \ref{fig3}a one notices that, qualitatively, the expected small-scale features are already reproduced in the first iteration step. However, the difference map in Fig.~\ref{fig3}c indicates, that the residual map has large-scale features that are misreconstructed. On the other hand, the corresponding difference map in Fig.~\ref{fig3}e after 20 iterations is closer to the expected map in Fig.~\ref{fig3}a.

The residual anisotropy map shown in Fig.~\ref{fig3}e is related to the Poisson variation of the event rate. We can make this statement more quantitative via a power spectrum analysis of expected and reconstructed anisotropy maps. The relative intensity can be decomposed as a sum over spherical harmonics, as in Eq.~(\ref{eq:Ylm}). Unfortunately, due to the limited integrated field of view, the true coefficients $\widehat{a}_{\ell m}$ cannot  be unambiguously reconstructed. However, for the present discussion of residual anisotropies in the iterative method, it is sufficient to study the pseudo multipole moments, ${a}_{\ell m}$, corresponding to the harmonic expansion of the anisotropy multiplied by the weight function of the field of view. In our example, the weight function is simply equal to $1$ for declinations $-41^\circ<\delta<79^\circ$ and $0$ otherwise. From these pseudo multipole moments, we can  compute the pseudo power spectrum,
\begin{equation}\label{eq:Cl}
C_\ell = \frac{1}{2\ell+1}\sum_{m}|a_{\ell m}|^2\,.
\end{equation}

\begin{figure}[t]
\centering
\includegraphics[width=0.95\linewidth]{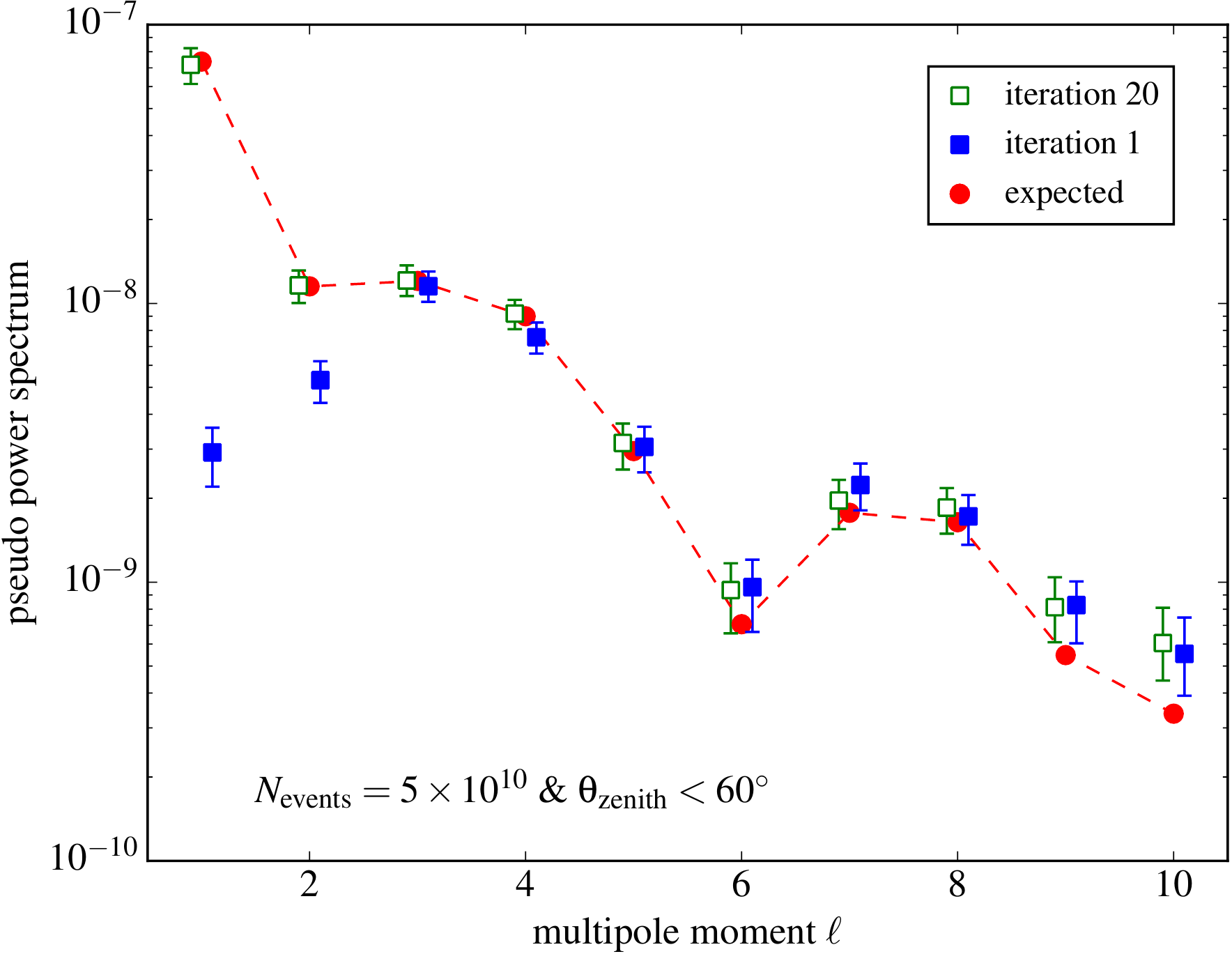}
\caption[]{Pseudo power spectra of reconstructed and expected anisotropy maps. The error bars on the power of the reconstructed anisotropy is indicating the variance from Poisson statistics estimated from 100 runs with the same anisotropy and detector model. Note that the small bias of the reconstructed pseudo power toward higher values for $\ell\geq6$ is due to the noise level of the maps as we will show in Section \ref{sec:harmonic}.}\label{fig4}
\end{figure}

Figure~\ref{fig4} shows the pseudo power spectrum of the anisotropy maps for the first and last iteration step in comparison to the expected spectrum corresponding to the true anisotropy with $\widehat{a}_{\ell0}=0$ and multiplied by the weight function. To estimate the variance of the reconstructed power spectra due to Poisson statistics, we repeat the analysis 100 times with low-resolution maps ($n_{\rm side}=16$ with bin size $\Delta\theta\simeq 4^\circ$) and show the central 68\% range of the data. One can see that the first iteration step drastically underestimates the power of the dipole ($\ell=1$) and quadrupole ($\ell=2$). This was already qualitatively visible in the map of Fig.~\ref{fig3}c. On the other hand, the pseudo power spectrum of the last iteration step agrees well with the low-$\ell$ multipoles. One can also notice a small bias of the $\ell\geq6$ multipoles towards larger values. This is due to the noise level of the maps as we will discuss in detail in Section \ref{sec:harmonic}.
 
\section{Comparison to Other Methods}\label{sec:comparison}

Various reconstruction methods have been developed previously to extract the anisotropy from the isotropic background. We have already mentioned the methods of direct integration~\cite{Atkins:2003ep} and time scrambling~\citep{Alexandreas1993} in the introduction, which have been used in cosmic-ray anisotropy studies of Super-Kamiokande~\citep{Guillian:2005wp}, Tibet-AS$\gamma$~\citep{Amenomori:2006bx}, Milagro~\citep{Abdo:2008kr}, IceCube~\citep{Abbasi:2011ai,Abbasi:2011zka,Aartsen:2013lla}, IceTop~\citep{Aartsen:2012ma}, HAWC~\citep{Abeysekara:2014sna}, and ARGO-YBJ~\citep{ARGO-YBJ:2013gya}. The time-integration method is closely related to the maximum-likelihood method presented here. The optimal relative detector acceptance (\ref{eq:Anull}) and the isotropic background expectation (\ref{eq:Nnull}) of the null hypothesis ($\delta I =0$) are equivalent to the estimates via direct integration with an integration period of $\Delta t=24$~h. The map of Fig.~\ref{fig3}b shows the resulting anisotropy estimate of this method.

Another two-dimensional reconstruction technique is the {\it Forward-Backward} method used by Milagro~\citep{Abdo:2008aw}, which is closely related to the one-dimensional {\it East-West} method~\citep{Bonino:2011nx}. These methods analyze the relative right-ascension derivative of event rates, $\partial_{\alpha}n/n$, at each sidereal time, either for individual declination bands (in the case of Forward-Backward) or the entire field of view (in the case of East-West). The anisotropy can then be reconstructed from the first derivative, noting that $\partial_{\alpha}\delta I \simeq\partial_{\alpha}n/n$,  up to an overall normalization constant in each declination band. This again reflects the invariance (\ref{eq:scaleI}) to (\ref{eq:scaleA}) and the inability to reconstruct the $a_{\ell 0}$ moments in equatorial coordinates. 

In general, the Forward-Backward (and the East-West) method has the advantage that the quantity $\partial_{\alpha}n/n$ does not explicitly depend on the instantaneous background level. For instance, the local deficit of background events at a local sidereal time of 9h in the example shown in Fig.~\ref{fig1} does not affect the estimate of the derivative $\partial_{\alpha}n/n$. However, the reconstruction methods discussed in \cite{Abdo:2008aw} assume that the relative detector acceptance is quasi-symmetric under Forward-Backward (or East-West) reflection, corresponding to the transformation $(\theta,\varphi) \to(\theta,-\varphi)$. This condition can be expressed in terms of the asymmetry $\epsilon \equiv [\mathcal{A}(\theta,\varphi)-\mathcal{A}(\theta,-\varphi)]/[\mathcal{A}(\theta,\varphi)+\mathcal{A}(\theta,-\varphi)]$ as $|\epsilon|\ll1$. After proper renormalization of the reconstructed derivative terms~\citep{Abdo:2008aw} the leading order effect is an $\mathcal{O}(\epsilon^2)$ correction of the anisotropy amplitude. We expect that the condition $|\epsilon|\ll1$ is met by most observatories. However, our method is also applicable for large  asymmetries approaching $\epsilon^2\simeq1$.

Iterative methods for the reconstruction of the anisotropy have also been developed for and applied to data of Tibet-AS$\gamma$~\citep{Amenomori:2005pn,Amenomori:2010yr,Amenomori:2012uda} and ARGO-YBJ~\citep{Bartoli:2015ysa}. The {\it Equi-Zenith Angle} method~\citep{Cui2003ICRC} estimates the isotropic cosmic-ray background at a celestial bin $\mathfrak{a}$ to be the average of those acceptance-corrected events that arrived from the same zenith angle band as the field of view wraps around the celestial equator. This corresponds to a generalization of the ansatz (\ref{eq:E}) to $\mathcal{E} = \sum_s E^s\mathcal{A}^s$, where the sum runs over the different equi-zenith-angle sectors with individual background rates $E^s$ and relative acceptance $\mathcal{A}^s$. It is straightforward to generalize our likelihood-based method with this ansatz (see Appendix~\ref{sec:appA}). The likelihood-based iteration method presented in this paper has the advantage that it can be derived from a firm statistical approach. If the relative detector acceptance can be regarded as stable in local sidereal time over the entire field of view, it provides an even simpler iterative reconstruction method.

\section{Harmonic Analysis}\label{sec:harmonic}

We now turn to the harmonic analysis of the anisotropy. We have already introduced the harmonic expansion of the relative intensity in Eq.~(\ref{eq:Ylm}). Of particular relevance for the theory of cosmic-ray diffusion is the strength of the dipole components ($\ell=1$), but note that TeV--PeV cosmic-ray data also show significant multipole moments at smaller angular scales, like quadrupole ($\ell=2$), octupole ($\ell=3$), {etc.} 

We have already shown in Section~\ref{sec:simulation} that traditional anisotropy methods can significantly underestimate the low-$\ell$ pseudo power spectrum of the data. This effect can be compensated by the iterative method presented in Section~\ref{sec:lratio}. However, this method will not compensate for the loss in power of the $m=0$ coefficients of the analysis. The observed spectrum is hence always a systematic underestimation of the true anisotropy. 

An additional uncertainty comes from the limited integrated field of view of most observatories. In the ideal case of a $4\pi$ sky coverage, the multipole moments ${a}_{\ell m}$ of the reconstructed anisotropy would carry all the information of the anisotropy (except $\widehat{a}_{\ell 0}$).  However, as already discussed earlier, the partial sky coverage of individual experiments does only allow reconstructing the pseudo multipole moments spectrum of the (reduced) anisotropy multiplied by the weight function.

The pseudo multipole moments ${a}_{\ell m}$ are related to the true multipole moments $\widehat{a}_{\ell m}$ via a linear transformation (see, {e.g.}, the review by \cite{Efstathiou:2003dj}) 
\begin{equation}\label{eq:Kmatrix}
{a}_{\ell m} = \sum_{\ell'm'}K_{\ell m\ell'm'}\widehat{a}_{\ell'm'}\,,
\end{equation}
where the coupling matrix ${\bf K}$ depends on the multipole spectrum $b_{\ell m}$ of the weight function of the field of view (see Appendix~\ref{sec:appB}). For an azimuthally symmetric weight function the strength of the mixing between moments ${a}_{\ell m}$ and $\widehat{a}_{\ell' m}$ is determined by the moments $b_{k0}$ with $|\ell-\ell'|\leq k\leq \ell+\ell'$. This mixing can be small for individual moments as pointed out by \cite{Denton:2014hfa}. However, it is important to emphasize that the full multipole spectrum cannot be unambiguously reconstructed from a partial sky coverage since the infinite-dimensional matrix ${\bf K}$ is not invertible.

\begin{figure}[t]
\centering
{\scriptsize(a)}\\
\includegraphics[width=\linewidth]{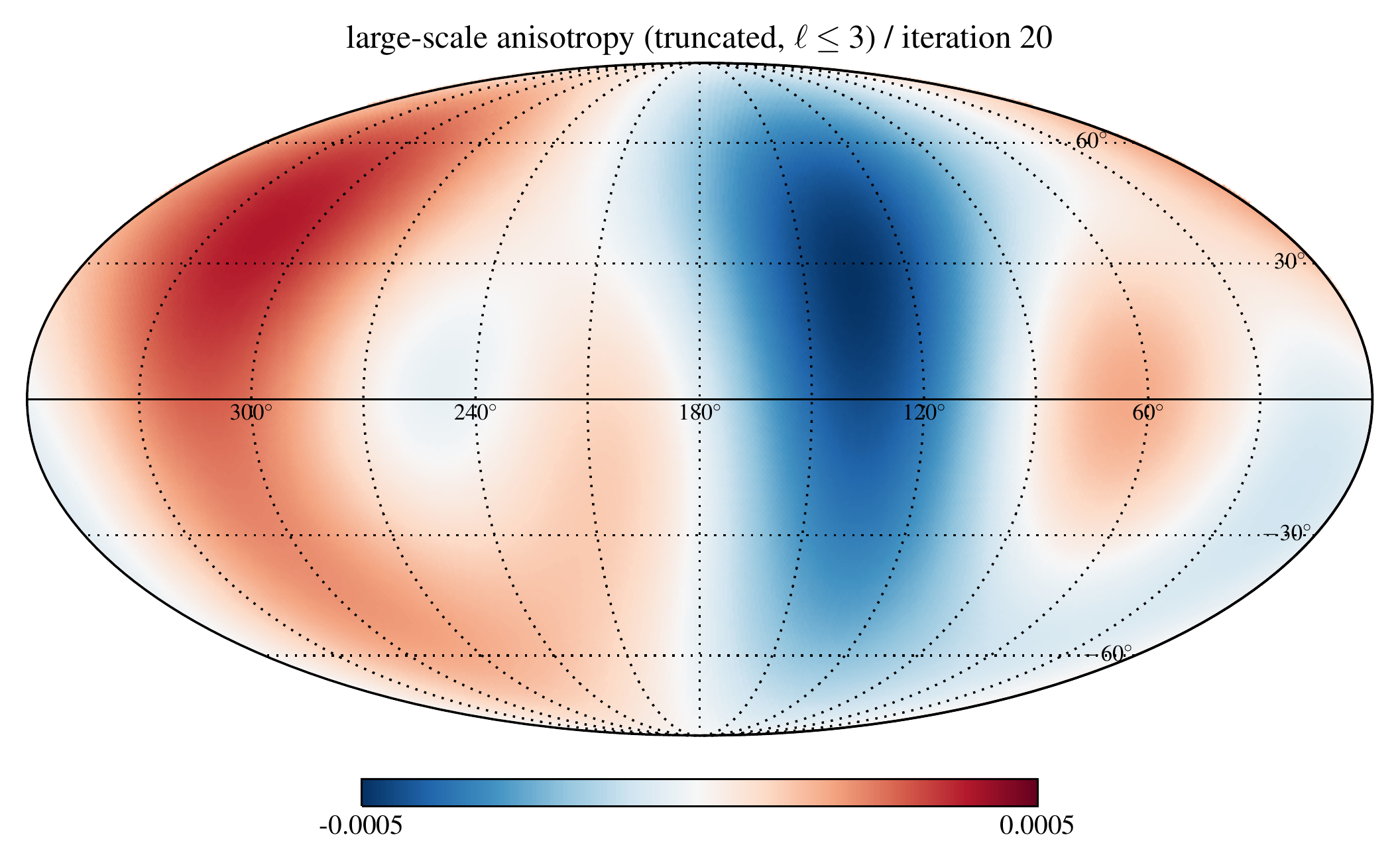}\\
{\scriptsize(b)}\\
\includegraphics[width=\linewidth]{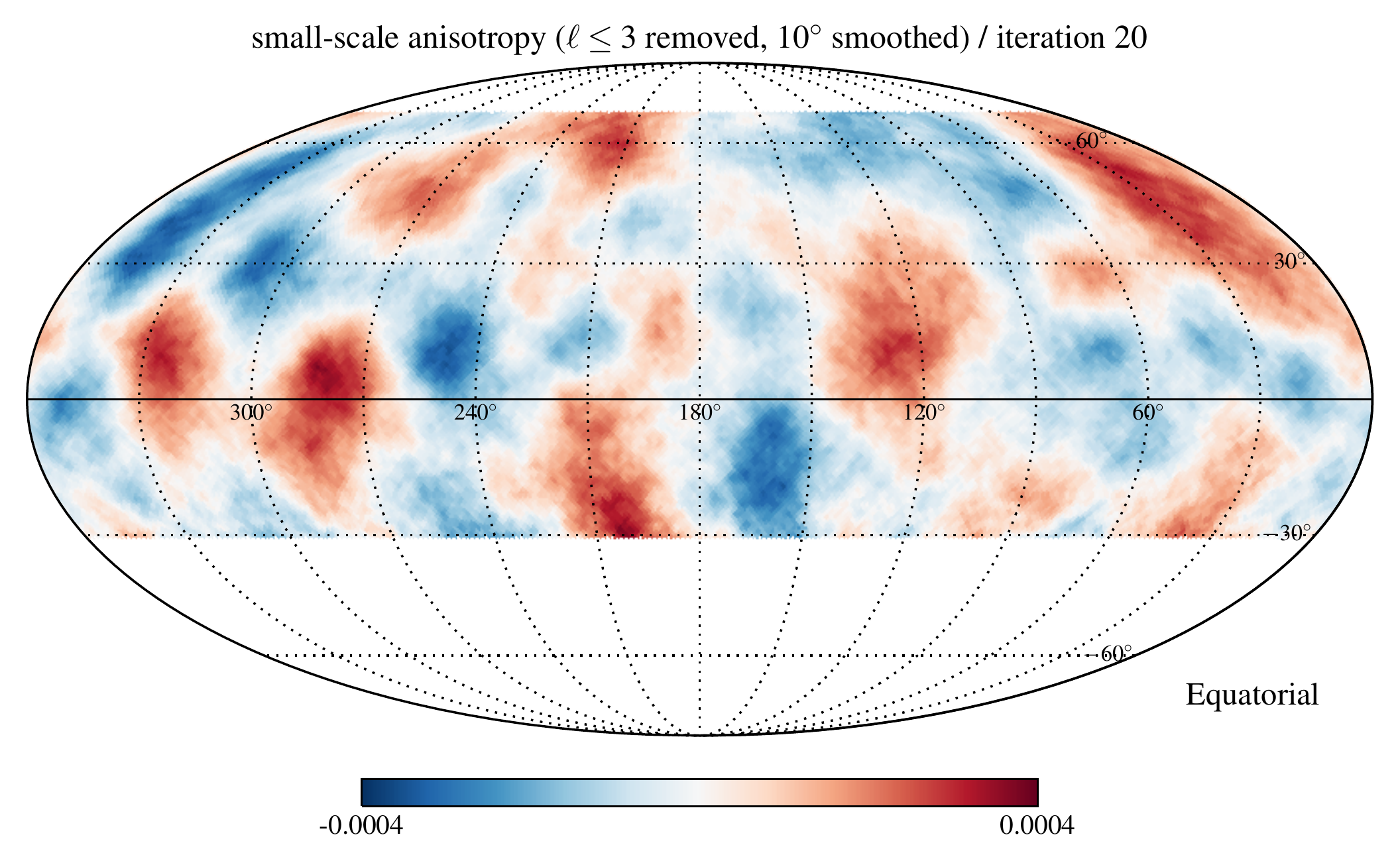}\\
{\scriptsize(c)}\\
\includegraphics[width=\linewidth]{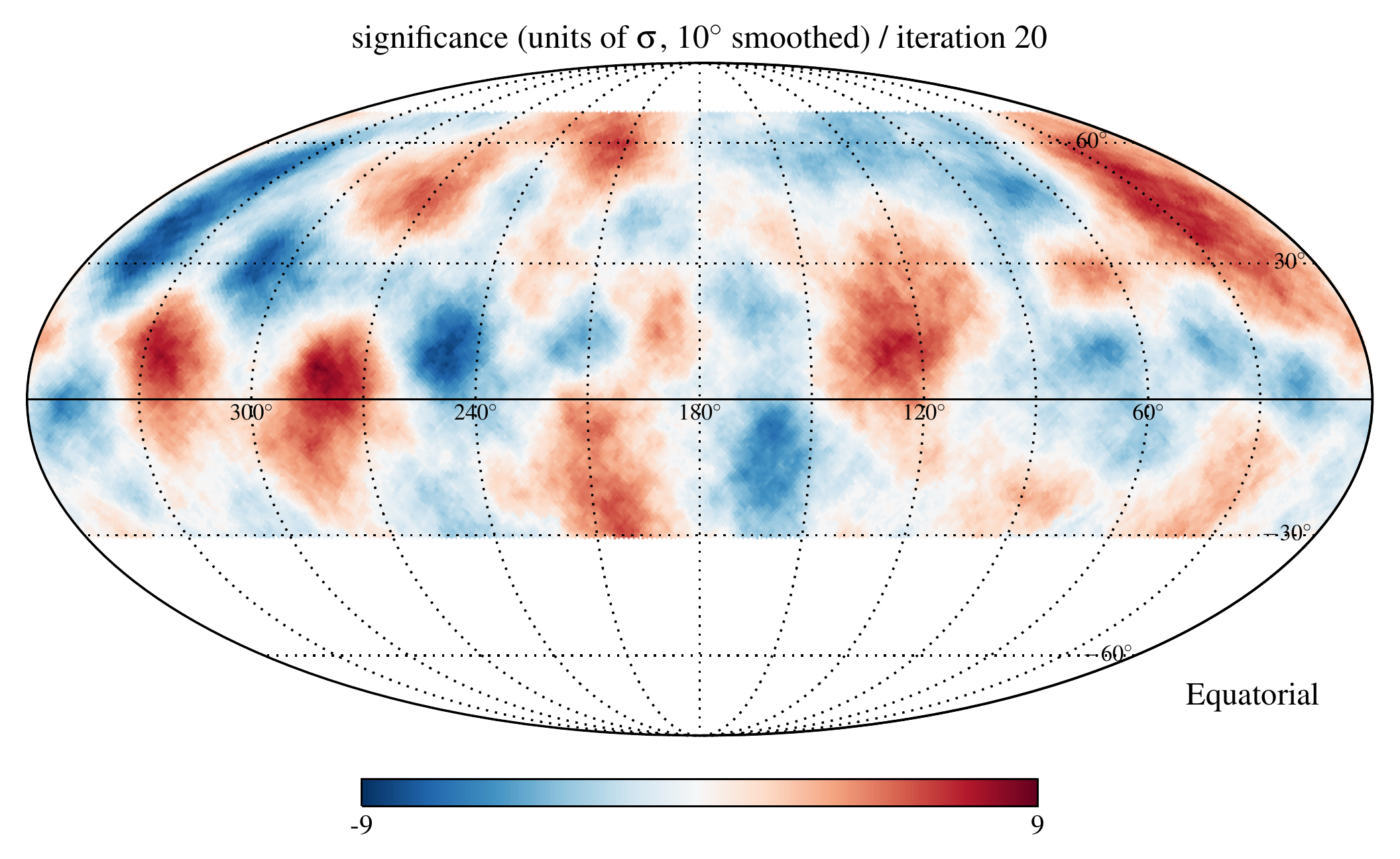}
\caption[]{{\bf Panel (a):} Reconstructed large-scale anisotropy from the solution of the matrix equation Eq.~(\ref{eq:Kmatrix}) truncated after $\ell=3$. {\bf Panel (b):} Residual small-scale anisotropy after subtracting the map in panel (a) from the full reconstructed anisotropy {\bf Panel (c):} Significance of the small-scale structure using Eq.~(\ref{eq:significance}). To distinguish excesses from deficits we multiply by the sign of the smoothed residual anisotropy $\delta I^{\rm small}$.}\label{fig5}
\end{figure}

\subsection{Large-Scale Anisotropy}

Whereas the full transition matrix ${\bf K}$ cannot be inverted, we can attempt an approximate reconstruction of the low-$\ell$ anisotropy via a truncation of the multipole expansion after a maximum $\ell$. The corresponding truncated matrix ${\bf K}'$ is then invertible. For instance, assuming a pure dipole anisotropy, $\ell\leq1$, and a uniform sky coverage between declination $\delta_1$ and $\delta_2$ gives the dipole transition elements
\begin{align}
K'_{1010} &= \frac{1}{2} \bigl(\sin^3\delta_2 -\sin^3\delta_1 \bigr)\,,\\
K'_{1111} &=\frac{1}{4} \bigl(3 (\sin\delta_2 -\sin \delta_1) 
   +\sin^3 \delta_1 - \sin^3 \delta_2\bigr) \,,
\end{align}
and $K'_{1\text{-}11\text{-}1} = K'_{1111}$. 

The dipole strength perpendicular to the Earth's rotation axis can then be estimated as
\begin{equation}
\widehat{A}_\perp = \sqrt{\frac{3}{4\pi}}\frac{\sqrt{|{a}_{11}|^2+|a_{1\text{-}1}|^2}}{K'_{1111}}\,.
\end{equation}
For instance, in our MC simulation with a zenith cut of $60^\circ$ and a latitude $\Phi\simeq19^\circ$, we have $\delta_1=-41^\circ$ and $\delta_2=79^\circ$ giving $K'_{1111}\simeq0.92$. Hence, the pseudo multipole moments of the projected dipole, $a_{11}$ and $a_{1\text{-}1}$, have to be corrected by a moderate factor $1/K'_{1111}\simeq1.09$ to recover the true moments $\hat{a}_{11}$ and $\hat{a}_{1\text{-}1}$. On the other hand, using the same zenith cut at the location of IceCube/IceTop gives $\delta_1=-90^\circ$ and $\delta_2=-30^\circ$, yielding $K'_{1111}\simeq0.16$ and a correction factor $1/K'_{1111}\simeq6.4$.

However, we emphasize that this treatment is only correct under the assumption that the true anisotropy is dominated by a dipole. In Fig.~\ref{fig5}a, we show a map of the reconstructed large-scale anisotropy including dipole ($\ell=1$), quadrupole ($\ell=2$), and octupole ($\ell=3$) with a truncation of the matrix in Eq.~(\ref{eq:Kmatrix}) after $\ell=3$. Note that the model of~\cite{Ahlers:2013ima} used for this simulation also assumes significant power at higher multipoles with $\ell>3$. The truncation after $\ell=3$ is hence not {\it per se} justified by this model. However, we can still use the reconstructed large-scale anisotropy map in Fig.~\ref{fig5}a as a background model to define the small-scale anisotropy in the full anisotropy map via subtraction. 

\subsection{Small-Scale Anisotropy}

The statistical significance of residual anisotropy features in the final reconstructed map is usually estimated using the method introduced by \citet{Li:1983fv} for applications in gamma-ray astronomy. A direct application of this method does not account for the optimization process of the time-dependent exposure. However, it is rather straightforward to generalize the method of \citet{Li:1983fv} to our case.

We begin by dividing the reconstructed relative intensity into a contribution of large-scale features and small-scale features, $I=I^{\rm large}+I^{\rm small}$. For each pixel $\mathfrak{a}$ in the celestial sky, we define expected {\it on-source} and {\it off-source} event counts in a disc of radius $\psi$ centered on that pixel.  Given the set of pixels $\mathcal{D}_\mathfrak{a}$, the observed and expected counts are
\begin{align}
  n_{\mathfrak{a}} &= \sum_{\mathfrak{b}\in\mathcal{D}_{\mathfrak{a}}}\sum_{\tau}n_{\tau \mathfrak{b}}\,,\\
  \mu_{\mathfrak{a},\rm on} &=  \sum_{\mathfrak{b}\in\mathcal{D}_{\mathfrak{a}}}\sum_{\tau}\mathcal{A}_{\tau\mathfrak{b}}\mathcal{N}_\tau I_{\mathfrak{b}}\,,\\
  \mu_{\mathfrak{a},\rm off} &= \sum_{\mathfrak{b}\in\mathcal{D}_{\mathfrak{a}}}\sum_{\tau}\mathcal{A}_{\tau\mathfrak{b}}\mathcal{N}_\tau I^{\rm large}_{\mathfrak{b}}\,.
\end{align}
The significance map (in units of Gaussian $\sigma$) is then calculated as
\begin{equation}\label{eq:significance}
  S_{\mathfrak{a}} = \sqrt{2}\left(-\mu_{\mathfrak{a},\rm on}+\mu_{\mathfrak{a},\rm off} + n_\mathfrak{a}\log\frac{\mu_{\mathfrak{a},\rm on}}{\mu_{\mathfrak{a},\rm off}}\right)^{1/2}\,.
\end{equation}
In Fig.~\ref{fig5}b, we show the residual anisotropy map after subtraction of the large-scale anisotropy map shown in Fig.~\ref{fig5}a. The corresponding significance map using (\ref{eq:significance}) is shown in Fig.~\ref{fig5}c. Usually, the significance is multiplied by the sign of the (smoothed) anisotropy (middle panel) to distinguish deficits (blue) from excesses (red), and we follow here the same convention. We can see that after the subtraction of large-scale features there is still significant power in small-scale features at a significance level of 8$\sigma$. 

Another approach to study small-scale structure is via the power spectrum (\ref{eq:Cl}), which quantifies the absolute amplitude of the multipole components but ignores their phases. We have already used the power spectrum to quantify the performance of our iterative method described in Section~\ref{sec:lratio}. We can also use this quantity in situations where the small-scale anisotropy structure is dominated by random scattering of cosmic rays in local magnetic fields~\citep{Giacinti:2011mz, Ahlers:2013ima, Ahlers:2015dwa, Lopez-Barquero:2015qpa}.

\begin{figure}[t]
\centering
\includegraphics[width=\linewidth]{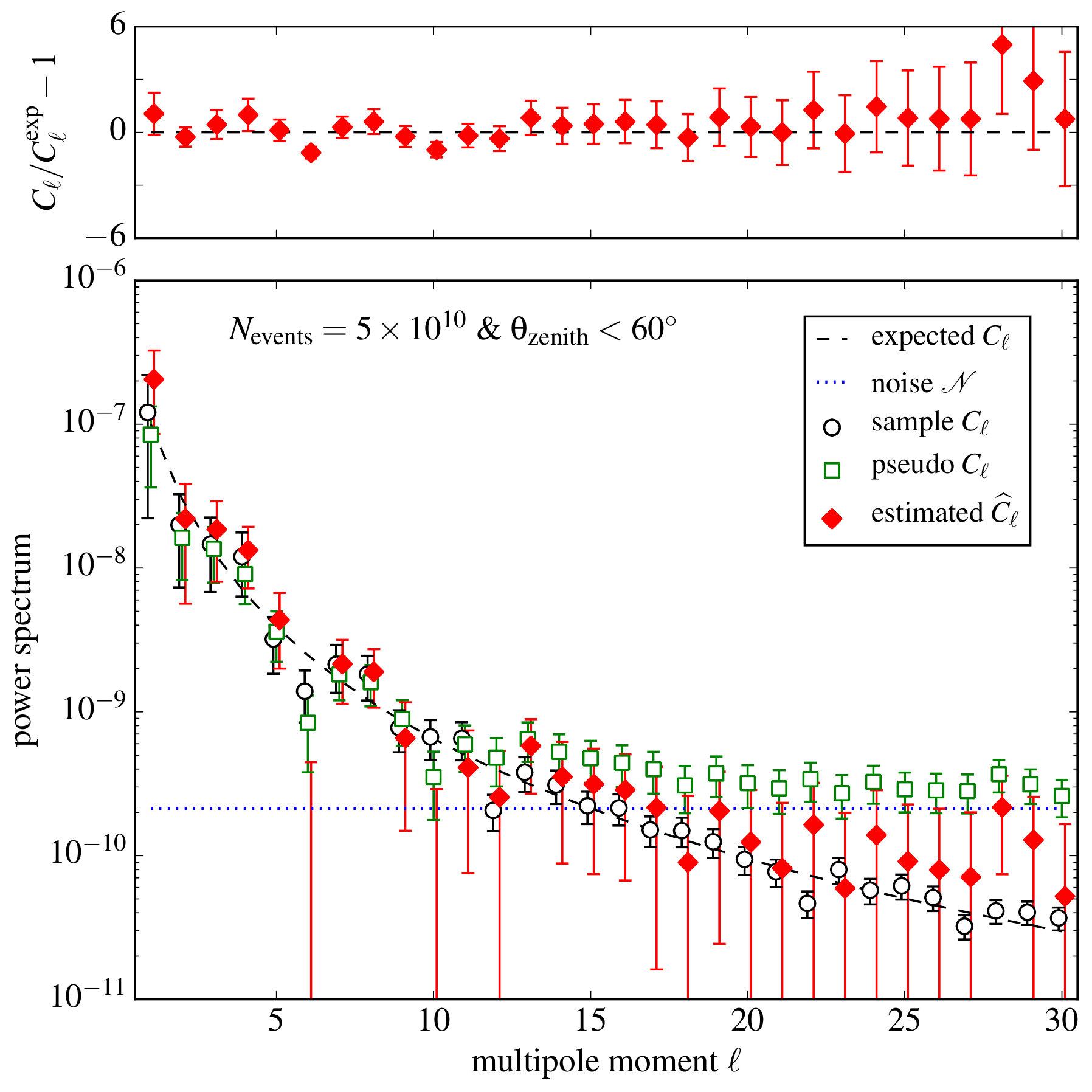}
\caption[]{Multipole power spectra of the final iteration map of the example. The dashed line shows the model expectation of the input power and the open circles show the sampled power spectrum of the input anisotropy map. The open boxes are the pseudo power spectrum for the reconstructed anisotropy. The filled diamonds show the estimated power spectrum accounting for the noise level (dotted line) and the field of view using the methods described in the main text. The upper plot shows the relative scatter of the reconstructed power spectrum from the expected input power.}\label{fig6}
\end{figure}

As mentioned in the previous section, the true multipole spectrum cannot be recovered unambiguously from a partial sky coverage, and the same is true for the power spectrum. However, in certain situations one can make additional assumptions about the {\it ensemble}-averaged expectation values of the multipole components. In the following, we will {\it assume} that the harmonic coefficients are Gaussian random fields, which in the ensemble-average follow $\langle\widehat{a}_{\ell m}\widehat{a}^*_{op}\rangle = \delta_{mp}\delta_{\ell o}\langle\widehat{C}_\ell\rangle$. In this particular case, we can recover the ensemble-averaged power spectrum $\langle\widehat{C}_\ell\rangle$ via the relation
\begin{equation}\label{eq:Mmatrix}
\langle C_\ell\rangle =\sum_{\ell'}M_{\ell\ell'}\langle \widehat{C}_{\ell'}\rangle +\mathcal{N}_\ell\,.
\end{equation}
The transfer matrix ${\bf M}$ is known from the study of temperature anisotropies in the cosmic microwave background~\citep{Efstathiou:2003dj}. However, for our situation of cosmic-ray anisotropies we have again to account for the fact that the $m=0$ moments are filtered out by the reconstruction. This leads to a modified expression for ${\bf M}$ that we discuss and provide in Appendix~\ref{sec:appB}.

In Eq.~(\ref{eq:Mmatrix}), we have also introduced the noise power spectrum $\mathcal{N}_\ell$, which can in general  be calculated from the relative intensity variance from the likelihood function (\ref{eq:LH}). In the following, we will use an approximation that only depends on pixel-by-pixel Poisson noise, which gives a flat spectrum $\mathcal{N}_\ell=\mathcal{N}$ with
\begin{equation}\label{eq:noise}
\mathcal{N} \simeq \frac{1}{4\pi}\sum_\mathfrak{a}\frac{w_\mathfrak{a}^2\Delta\Omega^2}{\sum_\tau n_{\tau\mathfrak{a}}}\,.
\end{equation}

In Fig.~\ref{fig6}, we show the estimated power spectrum inferred from the relative intensity map of the last iteration step shown in Fig.~\ref{fig3}d. The dashed line corresponds to the expected input power from which the input power spectrum (black data points) is sampled as a Gaussian random field. The sampling introduces a scatter of $(\Delta C_\ell)^2 = 2C_\ell^2/(2\ell+1)$ around the expected input power ({\it cosmic variance}). The green data points show the pseudo power spectrum, not accounting for the weight function. The horizontal blue dotted line shows the noise level of Eq.~(\ref{eq:noise}). Note that the noise level scales as $\mathcal{N}\propto1/N_{\rm tot}$ with the total number of events. The red data points are the best estimators of the true power spectrum $\langle\widehat{C}_\ell\rangle$. The variance of the pseudo and estimated power spectra are given in Appendix~\ref{sec:appB}.

\section{Summary}\label{sec:summary}

In this paper, we have discussed a novel two-dimensional cosmic-ray reconstruction method. It is based on a maximum likelihood analysis that provides implicit best-fit expressions for the relative intensity, relative detector acceptance, and background expectations. We have provided a detailed iterative method onhow the relative intensity can be reconstructed from these implicit maximum likelihood solutions. The performance of this likelihood-based method was studied via a simulated example, mimicking the position and performance of the HAWC observatory.

In general, ground-based observatories are insensitive to cosmic-ray anisotropy variations across declination bands. In terms of a spherical harmonic expansion of the anisotropy in equatorial coordinates this corresponds to a filter of $m=0$ multipole moments, which introduces a systematic underestimation of the observed anisotropy. In particular, the dipole anisotropy, which is a crucial observable for the study of cosmic ray diffusion in our Galaxy, can only be observed as a projection onto the celestial equator.

This has important consequences for the interpretation of experimental dipole data. If the dipole orientation changes with rigidity the projected dipole can exhibit rigidity modulations introduced by the projection on top of a {\it true} rigidity dependence of the dipole amplitude. If the dipole vector aligns with the celestial poles these projection effects can lead to a drastic reduction of the observed dipole accompanied by a phase-flip.

In addition, the limited integrated field of view of observatories affects the reconstruction of the multipole moments. As as consequence, even the $m=0$ filtered anisotropy cannot be unambiguously reconstructed. This can introduce a large systematic uncertainty, in particular, for the low-$\ell$ multipole moments of the anisotropy. We have discussed strategies how to account for this effect in the multipole reconstruction.

We would like to conclude with a few remarks. In our anisotropy ansatz (\ref{eq:phi}), we have not accounted for variations in the primary cosmic-ray flux during the observation period. The solar potential, with a variation on time scales of 11~years, only affects low-energy data and can be neglected for cosmic rays in the TeV to PeV energy range~\citep{Munakata:2009di}. On the other hand, the relative motion of the Earth in the solar system is expected to be visible as a solar dipole~\citep{Compton:1935}. This effect would be visible if the data is binned in terms of solar time. Indeed, a solar dipole anisotropy has been observed at the level of $10^{-4}$ in multi-TeV cosmic-ray data ~\citep{Amenomori:2004bf,Amenomori:2006bx,Abdo:2008aw,Abbasi:2011ai,Abbasi:2011zka,Bartoli:2015ysa}. However, the solar dipole is expected to average out for the data binned in sidereal time as long as the observation period covers an integer number of full years.

The method presented in this paper is designed for midlatitude observatories that are exposed to different parts of the celestial sky as the Earth rotates. The method is also applicable to the case of IceCube/IceTop located at the South Pole. However, for these observatories, the instantaneous field of view is identical to the time-integrated one. As a result, the iteration does not improve the estimates from  direct integration or time scrambling. On the other hand, the special location in combination with a limited field of view makes these observatories particularly insensitive to low-$\ell$ multipoles due to projection effects of the anisotropy discussed in the text.

In this paper, we have only considered the case of the data analysis of individual observatories. A great advantage of the likelihood-based reconstruction method is the straightforward generalization to {\it combined} anisotropy studies of data sets from multiple observatories with overlapping integrated field of view, see e.g.~\citep{DiazVelez2015}. The expectation value (\ref{eq:mu}) can then be simply generalized to a sum over data sets with individual detector exposures but same anisotropy, as long as the rigidity distributions of the data sets are very similar.

We would also like to emphasize that cosmic-ray observations via satellites like {\it Fermi} could have an advantage since the observatory can be tilted during the observation. In this case, it is possible to break the degeneracy between local acceptance effects and cosmic-ray anisotropy. In principle, it is then possible to have a full reconstruction of the anisotropy without projection effects, provided that additional systematic effects are under control.

\begin{acknowledgements}
{\it Acknowledgements.} The authors acknowledge support by the National Science Foundation (PHY-1306465, PHY-1306958, and PHY-1308033), by the Department of Energy (DE-SC0008475) and by the Wisconsin Alumni Research Foundation.
\end{acknowledgements}

\appendix

\section{Generalization of the Exposure Ansatz}\label{sec:appA}

We can generalize the ansatz~(\ref{eq:E}) by expressing the total accumulated exposure $\mathcal{E}$ as a sum over disjoint sky {\it sectors}, whose union covers the entire field of view. As before, we assume that the exposure in each sector can be expressed as a product of its angular-integrated exposure $E^s$ and relative acceptance in terms of azimuth $\varphi$ and zenith angle $\theta$ as
\begin{equation}\label{eq:Egen}
  \mathcal{E}(t,\varphi,\theta) \simeq \sum_{{\rm sector}\,{s}}E^{s}(t)\mathcal{A}^{s}(\varphi,\theta)\,.
\end{equation}
The partition of the field of view into multiple sectors is at this stage arbitrary and should in general be guided by the property of the data. For instance, the Equal-Zenith-Angle Method divides the sky into ring-like sectors with limited zenith angle range, {i.e.}, we have sector weight functions defined by $w^s(\theta) = 1$ if $\theta_s<\theta<\theta_{s+1}$ and otherwise $w^s(\theta) = 0$.  With this new ansatz, the best-fit relative acceptance and background rate of the null hypothesis become
\begin{equation}
  \mathcal{N}_\tau^{{s}\, (0)} =  {\sum_i w^{s}_in_{\tau i}}\,,\qquad
  {\mathcal{A}}_i^{{s}\, (0)} = \sum_\tau w^{s}_in_{\tau i}\Big/\sum_{\kappa j}w^{s}_jn_{\kappa j}\,.
\end{equation}
Here, we have introduced the weight function $w^s_i$ of the sector $s$ which is equal to $1$ if the pixel $i$ is located in the sector and $0$ otherwise. The maximum of the signal hypothesis now obeys the implicit relation
\begin{equation}
  {I}^\star_{\mathfrak{a}} =
  \sum_{\tau} n_{\tau\mathfrak{a}}\Big/ \sum_{{s}\kappa}\mathcal{A}^{s\,\star}_{\kappa \mathfrak{a}}\mathcal{N}^{{s}\,\star}_\kappa \,,\qquad
  \mathcal{N}^{{s}\,\star}_\tau =
  \sum_{i} w^{s}_i n_{\tau i}\Big/\sum_{j}\mathcal{A}^{s\,\star}_jI^\star_{\tau j}\,,\qquad
\mathcal{A}^{s\,\star}_i= \sum_\tau w^{s}_i n_{\tau i}\Big/\sum_{\kappa}\mathcal{N}^{s\,\star}_\kappa I^\star_{\kappa i}\,.\end{equation}

\section{Power Spectrum Estimator and Variance}\label{sec:appB}

In the following, we will assume that the weight function is azimuthally symmetric. In this case, the transfer function is block-diagonal $K_{\ell m\ell'm'} = \delta_{mm'}T^m_{\ell\ell'}( w)$ with block elements defined via a sum over Wigner-$3j$ coefficients,
\begin{equation}
T^m_{\ell\ell'}( b) = (-1)^{m}\sum_{k=|\ell-\ell'|}^{\ell+\ell'} b_{k0}\sqrt{\frac{(2\ell+1)(2\ell'+1)(2k+1)}{4\pi}}\begin{pmatrix}\ell&\ell'&k\\0&0&0\end{pmatrix}\begin{pmatrix}\ell&\ell'&k\\m&-m&0\end{pmatrix}\,.
\end{equation}
For the ensemble-averaged multipole moments, we can evaluate the transfer matrix to
\begin{equation}
M_{\ell\ell'} = \frac{2\ell'+1}{4\pi}\sum_{k}(2k+1)W_{k}\begin{pmatrix}\ell&\ell'&k\\0&0&0\end{pmatrix}^2 - \frac{[T^0_{\ell\ell'}( w)]^2}{2\ell+1}\,.
\end{equation}
Note that the unfamiliar last term in the previous equation accounts for the projection of the pseudo angular momentum onto $m\neq0$ terms.

The variance can be expressed as $\langle\Delta C_\ell \Delta C_{\ell'}\rangle = \mathcal{V}^1_{\ell\ell'}+\mathcal{V}^2_{\ell\ell'}+\mathcal{V}^3_{\ell\ell'}$ with
\begin{align}
\mathcal{V}^1_{\ell\ell'} &= \frac{2}{(2\ell+1)(2\ell'+1)}\sum_{m\neq0}\sum_{k}\sum_{k'}\langle \widehat{C}_{k}\rangle\langle\widehat{C}_{k'}\rangle  T^m_{k\ell}( w)T^m_{\ell'k}( w) T^m_{\ell k'}( w)T^m_{ k'\ell'}( w)\,,\\
\mathcal{V}^2_{\ell\ell'} &= \frac{2}{(2\ell+1)(2\ell'+1)}\sum_{m\neq0}\sum_{k}\langle \widehat{C}_{k}\rangle\left[  T^m_{\ell\ell'}( u)T^m_{k\ell}( w)T^m_{\ell'k}( w) +T^m_{\ell'\ell}( u)T^m_{k\ell'}( w)T^m_{\ell k}(w)\right]\,,\\
\mathcal{V}^3_{\ell\ell'} &= \frac{1}{2\pi}\sum_{k} (2k+1)U_{k}\begin{pmatrix}\ell&\ell'&k\\0&0&0\end{pmatrix}^2\,,
\end{align}
where $u_{\ell m}$ is the multipole coefficient of the distribution $\Delta\Omega w_\mathfrak{a}^2/\sum_\tau n_{\tau\mathfrak{a}}$ and $U_\ell$ the corresponding power spectrum. Since the variance matrix of $C_\ell$ is the same as for $C_\ell-\mathcal{N}$, we can express the variance of the true spectrum as 
\begin{equation}
\langle\Delta \widehat{C}_\ell \Delta \widehat{C}_{\ell'}\rangle = M^{-1}_{\ell k}M^{-1}_{\ell' k'}\langle\Delta C_k \Delta C_{k'}\rangle\,.
\end{equation}

\bibliographystyle{apj}
\bibliography{LHreco.bib}

\end{document}